\documentclass[10pt,preprint]{aastex}

\slugcomment{To Appear in ApJL}
\received{June 12, 2006}
\revised{June 30, 2006}
\accepted{July 13, 2006}

\makeatletter
\def\ale{\mathrel{\mathpalette\gl@align<}}
\def\age{\mathrel{\mathpalette\gl@align>}}
\def\gl@align#1#2{\lower.6ex\vbox{\baselineskip\z@skip\lineskip\z@
\ialign{$\m@th#1\hfil##\hfil$\crcr#2\crcr\sim\crcr}}}
\makeatletter
\shorttitle{Far-IR Bow-Shock Nebula of R Hya} 
\shortauthors{Ueta et al.}

\begin{document}
 
\title{Detection of a Far-Infrared Bow-Shock Nebula Around R Hya: \\
the First MIRIAD Results} 

\author{%
T.\ Ueta\altaffilmark{1,2},
A.\ K.\ Speck\altaffilmark{3},
R.\ E.\ Stencel\altaffilmark{4},
F.\ Herwig\altaffilmark{5},
R.\ D.\ Gehrz\altaffilmark{6},
R.\ Szczerba\altaffilmark{7},
H.\ Izumiura\altaffilmark{8},
A.\ A.\ Zijlstra\altaffilmark{9},
W.\ B.\ Latter\altaffilmark{10},
M.\ Matsuura\altaffilmark{11},
M.\ Meixner\altaffilmark{12},
M.\ Steffen\altaffilmark{13},
\&
M.\ Elitzur\altaffilmark{14}}

\altaffiltext{1}{%
NASA Ames Research Center/USRA SOFIA Office,
MS 211-3,
Moffett Field, CA 94035;
tueta@sofia.usra.edu}

\altaffiltext{2}{%
NRC Research Associate/NASA Postdoctoral Program Research Fellow}

\altaffiltext{3}{%
Dept.\ of Physics \& Astronomy, 
Univ.\ of Missouri, 
Columbia, MO 65211}

\altaffiltext{4}{%
Dept.\ of Physics \& Astronomy,
Univ.\ of Denver, 
Denver, CO 80208}

\altaffiltext{5}{%
Theoretical Astrophysics Group,
LANL,
Los Alamos, NM 87545}

\altaffiltext{6}{%
Dept.\ of Astronomy,
Univ.\ of Minnesota,
Minneapolis, MN 55455}

\altaffiltext{7}{%
N.\ Copernicus Astronomical Center, 
Rabia\'{n}ska 8, 
87-100, Toru\'{n}, Poland}

\altaffiltext{8}{%
Okayama Astrophysical Observatory, 
National Astronomical Observatory, %of Japan, 
Kamogata, Asakuchi, Okayama 719-0232, Japan}

\altaffiltext{9}{%
Univ.\ of Manchester, 
School of Physics \& Astronomy, 
PO Box 88,
Manchester M60 1QD, United Kingdom}

\altaffiltext{10}{%
%NASA Herschel Science Center, 
NHSC,
MC 100-22, 
%California Institute of Technology, 
Caltech,
Pasadena, CA 91125}

\altaffiltext{11}{%
National Astronomical Observatory, % of Japan, 
Mitaka, Tokyo, 181-8588, Japan}

\altaffiltext{12}{%
%Space Telescope Science Institute, 
STScI,
3700 San Martin Drive, 
Baltimore, MD 21218}

\altaffiltext{13}{%
Astrophysikalisches Institut Potsdam, 
%Astrophysical Institute Potsdom,
An der Sternwarte 16, 
14482, Potsdam, Germany}

\altaffiltext{14}{%
Physics \& Astronomy Dept., 
Univ.\ of Kentucky, 
Lexington, KY 40506}

\begin{abstract}
We present the first results of the MIRIAD (MIPS [Multiband Imaging Photometer
 for Spitzer] Infra-Red Imaging of AGB [asymptotic giant branch]
 Dustshells) project using the Spitzer Space Telescope.
The primary aim of the project is to probe the material distribution in
 the extended circumstellar envelopes (CSE) of evolved stars and recover
 the fossil record of their mass loss history. 
Hence, we must map the whole of the CSEs plus the surrounding sky for
 background subtraction, while avoiding the central star that is
 brighter than the detector saturation limit. 
With our unique mapping strategy, we have achieved better
 than one MJy ${\rm sr}^{-1}$ sensitivity in three hours of integration and
 successfully detected a faint ($< 5$ MJy sr $^{-1}$), extended
 ($\sim400\arcsec$) far-infrared nebula around the AGB star R Hya. 
Based on the parabolic structure of the nebula, the direction of the
 space motion of the star with respect to the nebula shape, and the
 presence of extended H$\alpha$ emission co-spatial to the nebula, we
 suggest that the detected far-IR nebula is due to a bow shock at the
 interface of the interstellar medium and the AGB wind of this moving star.
This is the first detection of the stellar-wind bow-shock interaction
 for an AGB star and exemplifies the potential of Spitzer as a tool to
 examine the detailed structure of extended far-IR nebulae around bright
 central sources.
\end{abstract}


\section{Introduction}

Evolved stars of intermediate initial mass ($\sim1$ to 8 M$_{\odot}$)
are major contributors to the interstellar medium (ISM).
However, the mechanism by which the ISM is enriched is not well
understood. 
The circumstellar envelopes (CSEs) of evolved stars, asymptotic giant
branch (AGB) and post-AGB stars in particular, contain the fossil
record of their mass loss history, and therefore, have the potential to
verify many aspects of stellar evolution. 
IRAS and ISO data indicated that parsec-sized dusty CSEs exist around
these infrared (IR) objects (e.g.\ Young, Phillips, \& Knapp 1993). 
Moreover, these large CSEs show evidence for mass-loss variations
that correlate with evolution of the star  (e.g.\ \citealt{s00}).

Previous observations lacked the sensitivity and spatial resolution to
investigate the full extent and structure of these large dusty CSEs. 
Using the unique sensitivity and mapping capabilities of the Multiband
Imaging Photometer for Spitzer (MIPS; \citealt{mips}) on-board the
Spitzer Space Telescope \citep{sst}, we have been conducting 
far-IR imaging observations to probe the material distribution in the
extended CSEs around several evolved stars.
Our maps are the deepest yet and have the most complete spatial 
coverage, allowing the most detailed study into the CSE structure to date. 

In this letter, we report the first results of the MIPS IR Imaging
of AGB Dustshells (MIRIAD) project (Program ID 20258) for the AGB 
star R Hya. 
This star is a Mira variable well-known for its decreasing period
(Zijlstra, Bedding, \& Mattei 2002) and the presence of a detached CSE
\citep{y93,hi97,h98}.
Below, we will describe the observations (\S 2) and data
reduction (\S 3), and present the results and discussion (\S 4).

\section{Observations}

\begin{figure}
 \begin{center}
 \includegraphics[width=\columnwidth]{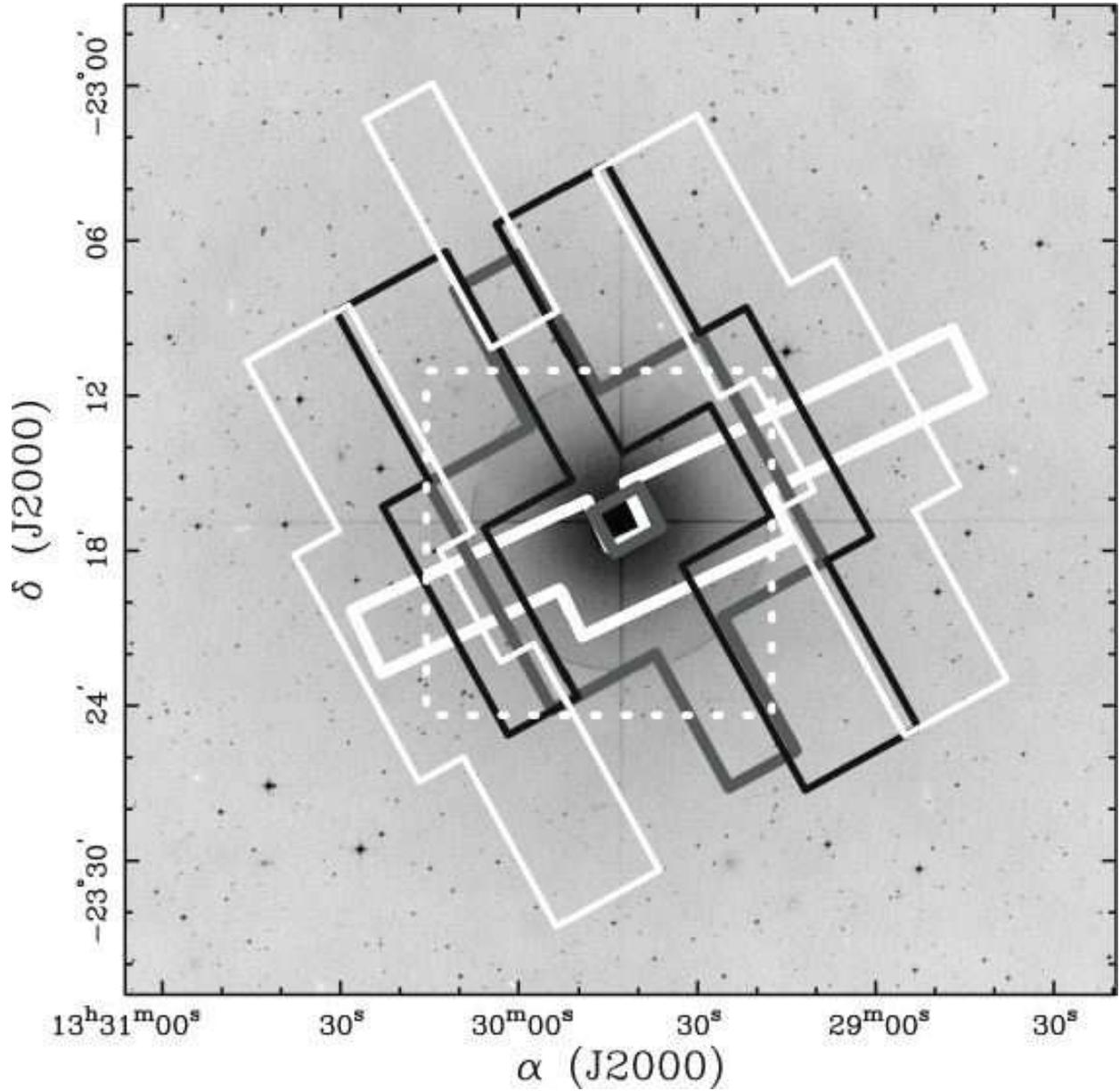}
 \end{center}
 \caption{\label{scheme}%
 MIRIAD sky coverage over R Hya: at $160\micron$ (thick
 white line) and at $70\micron$ for the outer (thin white line),
 intermediate (black line), and inner (gray line) regions.  
 The ``gap'' to the south of the star is unavoidable to prevent the
 central star from falling into the $24\micron$ array. 
 The central box (dashed white line) delineates the field of view of
 Fig.\ \ref{maps}.  
 The grayscale SERC-I plate image (at 790 nm) was taken from the
 Digitized Sky Surveys.}    
\end{figure}

We observed R Hya at 70 and $160\micron$ using Spitzer/MIPS on 2006
February 26 as part of the MIRIAD project. 
R Hya was the first target for which we obtained data in both of the
MIPS Ge bands. 
We mapped a roughly square region ($24\arcmin \times 24\arcmin$) at
$70\micron$ and a nearly linear region ($24\arcmin \times 2\farcm4$) at
$160\micron$, while avoiding the central region ($2\arcmin \times
1\arcmin$ and $1\farcm5 \times 2\arcmin$ at 70 and $160\micron$,
respectively), using exposures in photometry/fixed-cluster-offset mode. 
The MIRIAD maps can cover the entire target CSEs plus enough surrounding
sky for background subtraction. 
Care was taken to avoid the central star that is 
brighter than the saturation limit of the MIPS arrays. 
Fig.\ \ref{scheme} visualizes the sky coverage.

The $70\micron$ observations consisted of three Astronomical Observation
Requests (AORs), which sequentially covered the outer, mid, and inner
regions of the CSE.  
To maximize the dynamic range, we used progressively longer exposure
times for regions further away from the central star:
per-pixel, per-Data-Collection-Event (DCE) exposure times were 18.87,
62.91, and 251.66 s for the inner, mid, and outer regions,
respectively. 
The ``gap'' in the $70\micron$ sky coverage (along the in-scan direction
towards the south of the central star; Fig.\ \ref{scheme}) is
unavoidable because the central star would fall onto the $24\micron$
array when the $70\micron$ aperture covers this region.  

The $160\micron$ offsetting was done by a single AOR so that the linear
map was swept in both the forward and backward directions along the
cross-scan direction. 
To maximize the dynamic range, we repeated exposures at and near both
ends of the linear map: per-pixel exposure times are 167.76, 
251.56, and 335.52 s depending on the number of repeats.
Our observing scheme is also summarized in Table \ref{summary}.

\section{Data Reduction}

We initially constructed mosaicked maps from the pipeline-calibrated
basic calibrated data (BCD) products using the Mosaicker software (Ver.\
093005) provided by the Spitzer Science Center
(SSC)\footnote{\url{http://ssc.spitzer.caltech.edu/postbcd/}}.  
However, the BCD-mosaicked maps were affected by detector artifacts.
The effects were especially severe in the $70\micron$ map, in which
the ``streaking'' (due to residual variations of the slow response of
the detector as a 
function of time) was rampant and worsened 
increasingly towards the end of the observing sequence, probably
because residual cosmic-ray charge slowly built up in the detector. 

Hence, we performed a custom data reduction from the raw data.
First, we used the latest Ge Reprocessing Tools (GeRT) 
software (Ver.\ 041506 of S14
processing)\footnote{\url{http://ssc.spitzer.caltech.edu/mips/gert/}} 
to create our own BCDs. 
The most important step in custom BCD generation is to do proper
stimulator-flash (stim) calibration \citep{gordon}. 
For the data from the mid and outer AORs at $70\micron$,
for which most of the emission is of sky (the zodiacal light, ISM, and
cosmic IR background), the best results were 
produced from stim calibrations with piece-wise linear fitting
excluding stim frames on the nearby point sources.
For the data from the inner region at $70\micron$, for which most of the
field can be bright by the central star and CSE emission, 
the best outcome was yielded from calibrations with spline fitting
using all the raw data. 
For the $160\micron$ data, for which emission is of both sky and
non-sky, the best results were obtained from stim calibrations
with piece-wise linear fitting using all the raw data.  

For each BCD pixel, we followed the time evolution of pixel values   
and performed a linear least-squares fit using only the ``sky'' values
to determine the baseline for sky subtraction.
We tested this baseline fitting on the pixel, column, and
column-of-four-pixel (readout is done for four columns at a time) bases,
and the pixel basis fitting yielded the best results.
This process worked especially well in removing linear streakings in the
$70\micron$ maps which resulted from the column-dependent responsivity and
the relatively linear way in which photometry-mode mapping scans across the sky. 

However, the data obtained from the last offset position at $70\micron$
showed some residual column-dependent artifacts.
This probably occurred because the detector responsivity had been
compromised by the brightest region of the field near the central
star scanned immediately before the end of the
observing sequence (i.e.\ the detector is due for annealing).
The remaining artifacts were removed by flat-fielding all the
frames taken at the last offset position using their median frame.

After the baseline fitting, we subtracted the derived sky value at the
time of exposure from the entire dataset on pixel basis.
The subtracted sky values were 13.9, 14.4, and 15.5 MJy ${\rm sr}^{-1}$ 
respectively for the outer, mid and inner regions at
$70\micron$ and 19.0 MJy ${\rm sr}^{-1}$ at $160\micron$.
The increasing sky value with decreasing aperture at $70\micron$
illustrates the need of baseline determination far enough away from the
central object.
We adopted 13.9 MJy ${\rm sr}^{-1}$ as the $70\micron$ sky emission,
and the difference was added back to the BCDs for the mid and
inner regions (0.5 and 1.6 MJy sr$^{-1}$, respectively) to compensate
for the over-subtraction. 
The expected sky emission, obtained by the Spot
Software\footnote{http://ssc.spitzer.caltech.edu/propkit/spot/}, is 11.5
and 14.2 MJy ${\rm sr}^{-1}$ respectively at 70 and $160\micron$. 
This step was done by our own IDL script.

Finally, the custom-processed BCDs were mosaicked together using the
Mosaicker while cosmic-rays were removed.
The resulting maps still showed some structure intrinsic to the
diffraction spikes of the point-spread-function (PSF), especially at
$70\micron$.  
Thus, we used the Spitzer TinyTim (STinyTim)
software\footnote{\url{http://ssc.spitzer.caltech.edu/archanaly/contributed/}}
to create simulated point-response-function (PRF) maps (with the 
$F_{\nu} \propto \nu^{2}$ emissivity assumption), and subtracted
them from the mosaicked maps by scaling flux within an annulus around
the central unobserved region.

The resulting maps are in $4\farcs92$ pix$^{-1}$ and $7\farcs99$
pix$^{-1}$ (sub-pixelized from the nominal scale by a factor of
two) with the average of seven and 15 sky coverages per pixel and the
maximum of 45 and 80 sky coverages per pixel respectively at 
70 and $160\micron$. 
At least four sky coverages are achieved in $88\%$ and $94\%$ of the mapped
area respectively at 70 and $160\micron$ to fulfill the minimum level of
redundancy.  
The resulting one-sigma sensitivities are consistent with the
performance estimates done by the SENS-PET
tool\footnote{\url{http://ssc.spitzer.caltech.edu/tools/senspet/}}: we 
have achieved the required sensitivity level of 1 MJy ${\rm sr}^{-1}$ as
planned.  
Characteristics of the resulting maps are listed in Table \ref{perform}.

\section{Results and Discussion}

\begin{figure}
 \begin{center}
 \includegraphics[width=\textwidth]{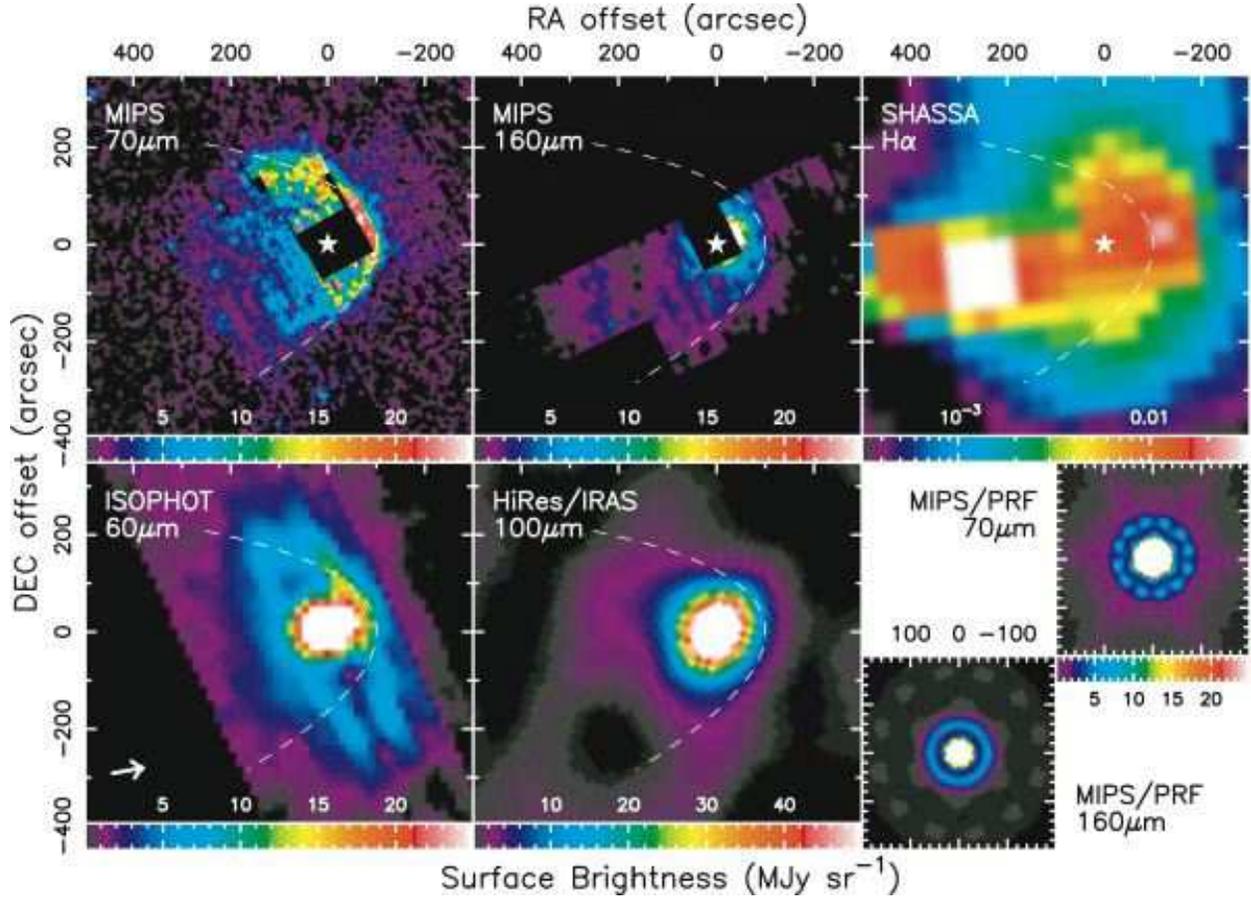}
 \end{center}
 \caption{\label{maps}%
 The background/PRF-subtracted, mosaicked MIPS maps of R Hya 
 (N is up, E to the left) at $70\micron$ (top left) and $160\micron$
 (top middle). 
 For comparison, an ISOPHOT PHT32 map at $60\micron$ (bottom left),
 a HiRes-processed IRAS map at $100\micron$ (bottom middle),
 a SHASSA H$\alpha$ map (top right), and 
 STinyTim PRF maps (bottom right) are also shown.  
 The images are zeroed at the position of R Hya (
 [$\alpha$, $\delta$]$_{2000}$ = [$13^{\rm h}29^{\rm m}42\fs7803$,
 $-23\arcdeg16\arcmin52\farcs792$]; \citealt{hipp}).
 The position of the star is indicated by the ``star''. 
 The tickmarks indicate the angular offsets in arcsec.  
 Linear color-scaling of surface brightness (in MJy sr$^{-1}$) is shown
 at the bottom of each panel.
 A parabolic curve $y = x^2 / 3 l$, which closely represents the
 stellar-wind bow-shock near the apex, is displayed by the dashed lines.
 The arrow at the bottom left corner shows the direction of the
 proper motion of the star, (-60.73, 11.01) mas yr$^{-1}$.} 
\end{figure}

Fig.\  \ref{maps} shows the background/PRF-subtracted, mosaicked 
MIPS color maps of R Hya at $70\micron$ (top left) and $160\micron$
(top middle), which clearly display faint, extended emission.
Also shown are 
a pipeline-processed ISOPHOT PHT32 map at $60\micron$ (bottom left;
\citealt{hi97})\footnote{\url{http://www.iso.vilspa.esa.es/}},
a HiRes-processed IRAS map at $100\micron$ (bottom
middle)\footnote{\url{http://irsa.ipac.caltech.edu/IRASdocs/hires\_over.html}},  
an H$\alpha$ map (top right) from the Southern H-Alpha Sky Survey
Atlas\footnote{\url{http://amundsen.swarthmore.edu/SHASSA/}}  
(SHASSA; \citealt{shassa})
and STinyTim PRF maps at 70 and $160\micron$ (bottom right).

The $70\micron$ map shows an arc-like surface brightness distribution
at $> 10$ MJy ${\rm sr}^{-1}$ surrounded by even fainter emission ($<
10$ MJy ${\rm sr}^{-1}$) of $\sim400\arcsec$ diameter. 
The arc structure appears parabolic with the apex pointed towards 
the WNW direction: the distance to the apex from the central star 
is $100\arcsec$ (see below).  
While the southern part of the arc remains physically thin
($\sim40\arcsec$), the northern part becomes wider ($\sim100\arcsec$). 
The width of the parabola at the tail end is about $370\arcsec$.
The brightest region near the apex of the parabolic arc (of $\sim20$
MJy ${\rm sr}^{-1}$) can also be seen in the ISOPHOT $60\micron$ map.
In retrospect, the ISOPHOT map appears to have captured most of the arc
to the north, in spite of the instrumental artifacts. 

The $160\micron$ map also displays an extended nebulosity along
the arc ($< 5$ MJy ${\rm sr}^{-1}$), which appears consistent 
with the $100\micron$ HiRes/IRAS map as well as the $100\micron$
high-resolution IRAS map based on the pyramid maximum entropy method
\citep{h98}.
There is a bright spot near the NW edge of the central unobserved
region (at $\sim45\arcsec$ from the star), which does not spatially
correspond to the $70\micron$ arc. 
It is probably caused by the near-IR light leak at
$160\micron$ (SSC 2005; For R Hya,  m$_{\rm J} = -1.3$ mag and
F$_{160\micron}$/F$_{2\micron} = 0.0006$).

The integrated flux over the entire nebula is $19.4\pm3.6$ and
$5.8\pm1.3$ Jy respectively at 70 and $160\micron$. 
These values are the lower limits since there is no data in the central 
unobserved region. 
From the scaled PRFs, the stellar fluxes at 70 and $160\micron$ are
estimated to be 33.9 and 4.5 Jy, respectively.

\citet{vbm88} have found parabolic far-IR nebulae around hot
stars and attributed some of them to bow shocks at the interface between
the stellar wind and ISM around a moving star.   
R Hya is known to have the proper motion of (-60.73, 11.01) mas
yr$^{-1}$ \citep{hipp}, towards the apex of the arc as indicated by the
bottom left arrow in Fig.\ \ref{maps}. 
Moreover, the shape of the far-IR arc closely follows the curve $y = x^2 
/ 3 l$ (where $l$ is the distance between the apex and star), which has
been shown numerically to represent the apex shape of the stellar-wind
bow-shock by \citet{mvbwc91}, as delineated by the 
dashed lines in Fig.\ \ref{maps}. 
Furthermore, H$\alpha$ emission is extended along the parabolic
apex and axis, showing the presence of ionized gas in these regions.

Therefore, we attribute the parabolic far-IR nebula around R Hya 
to shock-excited line emission (such as [\ion{O}{1}] 63 \& 146\micron)
and locally heated dust emission arising from
a stellar-wind bow-shock interface between the ISM and the swept-up AGB
wind of this moving star.
This is the first detection of the stellar-wind bow-shock interaction 
for an AGB star, and is a surprise discovery given our original aim of
the study. 
Nevertheless, our results exemplify the potential of Spitzer as a tool
to examine the far-IR structure of extended emission around the bright
central object. 
A follow-up IR spectroscopy can determine the emission
characteristics.

At 165 pc \citep{z02} with the -10 km s$^{-1}$ radial velocity
\citep{k98}, the space velocity of R Hya is $50\pm1$ km s$^{-1}$
into the direction 12$^{\circ}$ away (receding side) from the plane of 
the sky.  
The asymmetric CO profiles \citep{k98,t06} can then be explained by 
this inclined bow-shock, in which the receding and approaching shock
fronts intersect with the line of sight at distinct angles.

The physical dimensions of the observed bow shock (0.03 pc thick located 
at $7.9 \times 10^{-2}$ pc ahead of the star) are consistent with
the stellar-wind bow-shock models for AGB stars \citep{w06}.
Adopting the mass-loss rate of $3 \times 10^{-7}$ ${\rm M}_{\odot}$
${\rm yr}^{-1}$ \citep{z02}, wind velocity of 10 km s$^{-1}$ (e.g.,
\citealt{k98}), and formula for $l$ derived from 
the momentum conservation across the shock (eq.2 of \citealt{vbmwc90}),
the ambient H density is 0.4 cm$^{-3}$ and the shock surface density is
$2.5 \times 10^{-7}$ g cm$^{-2}$. 
Thus, the estimated amount of matter contained in the bow-shock is about
$1.3 \times 10^{-4}$ ${\rm M}_{\odot}$, assuming a paraboloidal
bow-shock of $y = x^2 / 3 l$ with a $300\arcsec$ opening radius. 

The MIPS maps indicate the presence of matter in the downstream
(emission at $\sim 3$ MJy ${\rm sr}^{-1}$ up to $300\arcsec$ away,
which is corroborated by the H$\alpha$ map).
The downstream emission may arise from a tail of a steady-state bow
shock that is formed by ram-pressure stripping from the head of the 
shock \citep{w06}.
The local emission peaks at $\sim250\arcsec$ away in the downstream may
correspond to vortices that results from instabilities in the downstream 
flow shed off from the bow shock (C.\ J.\ Wareing, in preparation). 
Modeling of the bow shock nebula of R Hya will be presented by
\citet{w06a}. 

\acknowledgements
This work is based on observations made with the Spitzer Space
Telescope, which is operated by the JPL/Caltech under a contract with
NASA.  
Support for this work was provided by NASA through an award issued by
JPL/Caltech.  
We also acknowledge additional support for the following individuals:
for Ueta by an NPP Research Fellowship Award, 
for Speck by NASA ADP grant (NAG 5-12675),
for Herwig by the LDRD program (20060357ER) at LANL,
for Gehrz by NASA (Contract 1215746) issued by JPL/Caltech,
for Szczerba by grant 2.P03D.017.25,
for Izumiura by Grant-in-Aid (C) from JSPS (No.17540221),
for Matsuura by JSPS,
and
for Elitzur by NSF AST grant (0507421).
We thank C.\ J.\ Wareing for sharing his insights on the 
stellar-wind bow-shocks with us.

\clearpage

\begin{deluxetable}{lcccccccccc}
%\rotate
\tablecolumns{11} 
\tablewidth{0pt} 
%\tablewidth{\textwidth} 
\tabletypesize{\tiny}
\tablecaption{\label{summary}%
Summary of Spitzer/MIRIAD Observations of R Hya} 
\tablehead{%
\colhead{} &
\colhead{} &
\colhead{} &
\colhead{Image Scale} &
\colhead{Field} &
\colhead{Exposure} &
\colhead{} &
\colhead{Integration\tablenotemark{a}} &
\colhead{} &
\colhead{DCE} &
\colhead{Total Duration} \\
\colhead{Band} &
\colhead{AOR KEY} &
\colhead{Start Time} &
\colhead{(arcsec)} &
\colhead{Size} &
\colhead{(sec)} &
\colhead{Cycle} &
\colhead{(sec)} &
\colhead{Offsets} &
\colhead{Counts} &
\colhead{(sec)}} 
\startdata 
$70\micron$ (outer) &
14472448 &
2006 Feb 26 07:15:03 &
\phn9.84 &
Large &
10 &
\phn2 &
125.83 &
13 &
418 &
\phn6493 \\

$70\micron$ (mid) &
14472704 &
2006 Feb 26 08:59:52 &
\phn9.84 &
Large &
10 &
\phn1 &
\phn62.91 &
\phn8 &
145 &
\phn2625 \\

$70\micron$ (inner) &
14472960 &
2006 Feb 26 09:40:12 &
\phn9.84 &
Large &
\phn3 &
\phn1 &
\phn18.87 &
\phn6 &
109 &
\phn1204 \\

$160\micron$  &
14472192 &
2006 Feb 26 04:20:36 &
16.0\phn &
Small &
10 &
\phn2 &
%\phn41.94 &
\phn83.88 &
22 &
792 &
10256 
\enddata

\tablenotetext{a}{Integration times are per-pixel per-DCE in ``real''
 second.}
\end{deluxetable} 

\clearpage

\begin{deluxetable}{lcccccccccc}
%\rotate
\tablecolumns{11} 
\tablewidth{0pt} 
\tablewidth{\columnwidth} 
\tabletypesize{\tiny}
\tablecaption{\label{perform}%
Summary of Measurements and Map Characteristics} 
\tablehead{%
\colhead{} &
\multicolumn{2}{c}{Flux (Jy)} &
\colhead{} &
\multicolumn{4}{c}{Sky, $\sigma$ (MJy sr$^{-1}$)} &
\colhead{} &
\multicolumn{2}{c}{Coverage (pix$^{-1}$)} \\
\cline{2-3}
\cline{5-8}
\cline{10-11}
\colhead{Band} &
\colhead{CSE\tablenotemark{a}} &
\colhead{Star\tablenotemark{b}} &
\colhead{} &
\colhead{Sky} &
\colhead{$\sigma_{\rm avg}$} &
\colhead{$\sigma_{\rm best}$} &
\colhead{$\sigma_{\rm est}$\tablenotemark{c}} &
\colhead{} &
\colhead{Avg} &
\colhead{Best}} 
\startdata 
$70\micron$ (mosaic) &
$19.4 \pm 3.6$ &
33.9 &
 &
13.9 &
0.47 &
0.15 &
\dots &
 &
\phn7 &
45 \\

$70\micron$ (outer) &
\dots &
\dots &
 &
13.9 &
0.56 &
0.16 &
0.26 &
 &
\phn5 &
45 \\

$70\micron$ (mid) &
\dots &
\dots &
 &
14.4 &
0.68 &
0.21 &
0.37 &
 &
\phn3 &
24 \\

$70\micron$ (inner) &
\dots &
\dots &
 &
15.5 &
0.86 &
0.28 &
0.68 &
 &
\phn2 &
17 \\

$160\micron$  &
$\phn5.8 \pm 1.3$ &
\phn4.5 &
 &
19.0 &
0.31 &
0.12 &
0.38 &
 &
15 &
80 
\enddata

\tablenotetext{a}{{L}ower-limit.}
\tablenotetext{b}{Based on the scaled PSF/PRF maps.}
\tablenotetext{c}{One-sigma sensitivity estimated with the SENS-PET tool.}
\end{deluxetable} 

\clearpage

\clearpage

\end{document}